\documentclass[conference]{IEEEtran}
\usepackage{amsmath,amssymb}
\usepackage{booktabs}
\usepackage{graphicx}
\usepackage{url}

\begin{document}

\title{Fast and Robust On-Device Speaker Diarization:\\
Relative Minimum Cluster Size for Stride-Accelerated Pipelines}

\author{\IEEEauthorblockN{Fumiaki Yamaguchi}
\IEEEauthorblockA{Independent Researcher\\
panchorange2203@gmail.com}}

\maketitle

\begin{abstract}
Speech applications such as meeting transcription and voice agents would
benefit from on-device speaker diarization, but practical adoption is limited
by inference cost. We study how far a Pyannote~3.1-based pipeline can be
accelerated on consumer hardware (an RTX~5070~Ti GPU and an Apple~M4 laptop)
while preserving diarization error rate (DER). A simple recipe---coarser
segmentation stride and per-chunk embedding---yields multi-fold speedups and is
DER-neutral on AMI, but degrades sharply on in-the-wild data: on VoxConverse,
DER rises from 0.075 to 0.113. We trace the failure to speaker under-counting in
the clustering stage, caused by a fixed minimum cluster size interacting with
the reduced number of embeddings per speaker. We propose a \emph{relative
minimum cluster size}, $\mathrm{mcs}=\mathrm{round}(f\cdot n)$ with $f\approx0.01$,
which adapts to the embedding budget per recording. A single value of $f$
recovers VoxConverse DER to 0.079 (about 89\% of the lost accuracy) while keeping
AMI flat, and the accelerated pipeline reaches up to $12.2\times$ speedup on AMI
(MPS) over our CAM++ baseline.
\end{abstract}

\begin{IEEEkeywords}
speaker diarization, inference efficiency, on-device, clustering
\end{IEEEkeywords}

\section{Introduction}

Speech recognition applications---meeting minutes, voice dialogue, automatic
scribes---are increasingly common, but their adoption is constrained by
performance. Cloud APIs cost money and require sending audio over the network,
which some users cannot accept; running heavy models locally requires technical
expertise and hardware that many users lack. This tension motivates a broader
shift toward on-device AI, where lightweight local inference is increasingly
seen as essential for data privacy and real-time responsiveness under tight
resource budgets~\cite{ondevice}. Reducing inference cost, expressed
as the real-time factor (RTF), is therefore central to bringing these systems to
everyday devices. Moreover, the magnitude of the speedup changes the interaction
model qualitatively, not merely quantitatively: processing an hour of audio in a
few minutes lets a user start the job and return later, whereas finishing in
seconds lets them wait for the result in place.

We focus on speaker diarization---the task of determining ``who spoke when.''
Modern systems such as Pyannote~\cite{pyannote} adopt a multi-stage pipeline:
(i)~\emph{segmentation} divides local windows into speaker-homogeneous segments,
(ii)~\emph{embedding} extracts a speaker representation per segment, and
(iii)~\emph{clustering} groups segments by speaker identity. This modular
structure invites stage-targeted optimization.

Recent work on inference-efficient diarization, such as SDBench and its
SpeakerKit system~\cite{sdbench}, reports large speedups from coarsening the
segmentation stride on top of Pyannote, but also observes that an aggressive
stride degrades accuracy, and that this degradation \emph{grows with the number
of speakers} (their Fig.~3: DER impacted by ${<}0.02$ for up to five speakers
but by ${<}0.05$ for unlimited speakers). They report this as a speed--accuracy
trade-off without resolving it. Our experiments reproduce the degradation, trace
it to speaker under-counting in the clustering stage---which explains the
speaker-count dependence---and recover the accuracy by adapting the clustering
granularity rather than backing off the stride.

In this work we perform performance engineering on a Pyannote~3.1-based pipeline
and ask how the resulting speedups change the user experience on consumer
hardware (an RTX~5070~Ti and an Apple~M4 laptop). Our contributions are:
\begin{itemize}
\item A speedup recipe (coarse stride + per-chunk embedding) and a demonstration
that its naive application breaks on in-the-wild data
(Section~\ref{sec:recipe}).
\item A diagnosis attributing the failure to speaker under-counting in the
clustering stage (Section~\ref{sec:diagnosis}).
\item A \emph{relative minimum cluster size} that resolves the failure with a
single hyperparameter across AMI, VoxConverse, and MSDWild
(Section~\ref{sec:relative}).
\end{itemize}

\section{Method}
\label{sec:method}

\subsection{Pipeline and evaluation conditions}
Unless otherwise noted, all systems share the segmentation model
\texttt{pyannote/\allowbreak segmentation-3.0}~\cite{powerset}, the
CAM++~\cite{campplus} embedding model
(\texttt{wespeaker-\allowbreak voxceleb-\allowbreak campplus}~\cite{wespeaker}, float16), and
agglomerative clustering (AHC, centroid linkage~\cite{ahc}). We evaluate on three datasets: AMI~\cite{ami}
(headset mix, test), VoxConverse~\cite{voxconverse} (test), and
MSDWild~\cite{msdwild} (many.val). We report accuracy as the diarization error
rate (DER), the sum of missed detection, false alarm, and speaker confusion
divided by the total reference speech duration, computed with
pyannote.metrics~\cite{pyannotemetrics}. DER uses a collar of
0.25\,s and with overlapped speech excluded (\texttt{skip\_overlap=true}),
following the AMI \emph{only\_words} evaluation convention also used by prior
work. We report speed as the real-time factor (RTF), the processing time divided
by the audio duration (equivalently, the inverse of the speed factor used
in~\cite{sdbench}); lower is faster, and RTF${}<{}$1 means faster than real
time. RTF is reported with the device made explicit, since it is
hardware-dependent: ``MPS'' denotes an Apple~M4 laptop, ``CPU'' the same laptop
restricted to its CPU, and ``CUDA'' an RTX~5070~Ti GPU. AMI RTF figures are measured on MPS;
VoxConverse RTF figures (Tables~\ref{tab:main} and~\ref{tab:crossdevice}) and MSDWild baseline RTF
(Table~\ref{tab:msdwild}) are measured on CUDA. Each RTF entry is tagged with its device, and DER is device-independent.

\subsection{Speedup recipe}
We increase the segmentation stride (the hop of the sliding window) from 1\,s to
3\,s and adopt per-chunk embedding (one embedding per chunk rather than per
frame). Both reduce the number of embeddings to be computed and clustered.

\subsection{Relative minimum cluster size}
AHC discards clusters smaller than a minimum cluster size (mcs). Instead of a
fixed value (e.g.\ $\mathrm{mcs}=12$), we set
\begin{equation}
\mathrm{mcs} = \mathrm{round}(f \cdot n),
\end{equation}
where $n$ is the total number of embeddings and $f$ is a single fraction. This
assigns a large mcs to recordings with abundant embeddings (e.g.\ AMI) and a
small mcs to recordings where embeddings are scarce (e.g.\ VoxConverse).

\section{A Speedup Recipe and How It Breaks}
\label{sec:recipe}

\subsection{Building the baseline}
Starting from a Pyannote~3.1 reimplementation on AMI test (16 files), we apply
two standard inference optimizations and an embedding-model swap
(Table~\ref{tab:baseline}). float16 and a larger batch reduce RTF without
changing DER; switching the embedding model to CAM++ further reduces RTF, at the
cost of a DER increase below 0.01, which we accept. The resulting configuration
(RTF~$0.061$ on MPS, i.e.\ processing one hour of audio in about 3.7~minutes) is
our baseline for the rest of the paper.

\begin{table}[t]
\centering
\caption{Baseline construction on AMI test (MPS).}
\label{tab:baseline}
\setlength{\tabcolsep}{4pt}
\begin{tabular}{lccc}
\toprule
Configuration & DER micro/macro & RTF & $\Delta$DER \\
\midrule
self-implement (ResNet34~\cite{resnet}) & 0.081 / 0.084 & 0.107 & --- \\
\;+ fp16 + bs64 & 0.081 / 0.084 & 0.074 & 0 \\
\;+ CAM++ embedding & 0.081 / 0.085 & \textbf{0.061} & $+{<}0.01$ \\
\bottomrule
\end{tabular}
\end{table}

\subsection{The recipe works on AMI but breaks on VoxConverse}
On AMI, both knobs reduce RTF monotonically while DER stays within 0.01 of the
baseline (Table~\ref{tab:ablation}, all on MPS). Combined, stride-3 plus
per-chunk embedding accelerates AMI by $9.9\times$ (RTF $0.061\!\to\!0.006$), and
adding relative mcs (Section~\ref{sec:relative}) reaches $12.2\times$
($\to0.005$). All speedup factors in this paper are AMI/MPS, measured against our
CAM++ baseline on the same device.

\begin{table}[t]
\centering
\caption{AMI speedup ablation (all on MPS, same pipeline).}
\label{tab:ablation}
\setlength{\tabcolsep}{4pt}
\begin{tabular}{lccc}
\toprule
Configuration & DER micro & RTF & Speedup \\
\midrule
CAM++ baseline (stride1) & 0.081 & 0.061 & 1.0$\times$ \\
\;+ per-chunk & 0.084 & 0.028 & 2.2$\times$ \\
stride2 & 0.080 & 0.025 & 2.4$\times$ \\
stride3 & 0.082 & 0.015 & 4.1$\times$ \\
stride3 + per-chunk & 0.082 & 0.006 & 9.9$\times$ \\
\;+ relative mcs ($f{=}0.01$) & 0.083 & \textbf{0.005} & \textbf{12.2$\times$} \\
\bottomrule
\end{tabular}
\end{table}

The speedup is not specific to the MPS backend. On the same Apple~M4 laptop
restricted to its CPU---a setting with neither a discrete GPU nor MPS, closer to
a typical consumer machine---coarsening the stride from 1 to 3\,s (with per-chunk
embedding) gives an $8.5\times$ speedup, cutting AMI RTF from $1.84$ to $0.22$ at
an unchanged DER (Table~\ref{tab:cpu}). An RTF of $0.22$ means about 13~minutes
of processing per hour of audio: still a wait, but firmly in the practical range
on hardware with no accelerator at all.

\begin{table}[t]
\centering
\caption{CPU-only stride speedup on AMI test (Apple~M4, no GPU/MPS).}
\label{tab:cpu}
\setlength{\tabcolsep}{4pt}
\begin{tabular}{lccc}
\toprule
Configuration & DER micro & RTF (CPU) & Speedup \\
\midrule
stride1 + per-chunk & 0.084 & 1.84 & 1.0$\times$ \\
stride3 + per-chunk & 0.083 & \textbf{0.22} & \textbf{8.5$\times$} \\
\bottomrule
\end{tabular}
\end{table}

On VoxConverse, however, the same recipe raises DER from $0.075/0.086$ to
$0.113/0.124$ (Table~\ref{tab:main}). The optimization is data-dependent and
fails on in-the-wild audio.

\begin{table*}[t]
\centering
\caption{AMI vs.\ VoxConverse: baseline $\to$ stride3+per-chunk $\to$ relative mcs.}
\label{tab:main}
\setlength{\tabcolsep}{5pt}
\begin{tabular}{lccccc}
\toprule
Configuration & AMI DER (micro/macro) & VoxConverse DER (micro/macro) & AMI RTF (MPS) & VoxConverse RTF (CUDA) \\
\midrule
baseline (stride1, mcs12) & 0.081 / 0.085 & 0.075 / 0.086 & 0.061 & 0.004 \\
stride3 + per-chunk, mcs12 & 0.082 / 0.088 & \textbf{0.113 / 0.124} & 0.006 & 0.001 \\
\;+ relative mcs ($f{=}0.01$) & 0.083 / 0.088 & \textbf{0.079 / 0.086} & 0.005 & 0.00083 \\
\bottomrule
\end{tabular}
\\[2pt]
\footnotesize VoxConverse RTF is all CUDA (RTX~5070~Ti); AMI RTF is MPS (Apple~M4). DER is device-independent.
\end{table*}

\section{Diagnosis}
\label{sec:diagnosis}

Decomposing the VoxConverse degradation shows that almost all of it is the
\emph{confusion} component ($+0.037$), while missed detection and false alarm are
essentially unchanged. The problem is therefore in clustering, not segmentation.

A coarser stride reduces the number of embeddings per speaker by roughly a factor
of three. VoxConverse has only about 10 embeddings per speaker, versus about 200
on AMI---a $\sim$20$\times$ difference. With a fixed $\mathrm{mcs}=12$, small
speakers in VoxConverse fall below the threshold and are absorbed into other
clusters, producing speaker under-counting (the predicted speaker count drops
below the reference). These absorbed speakers disappear from the output
entirely, and their speech is reassigned to the surviving clusters; this
misattribution is exactly what the confusion term measures, which is why the
degradation is almost purely confusion rather than missed detection. AMI, with
its abundant embeddings, is insensitive to mcs and is unaffected.

\begin{figure*}[t]
\centering
\includegraphics[width=\textwidth]{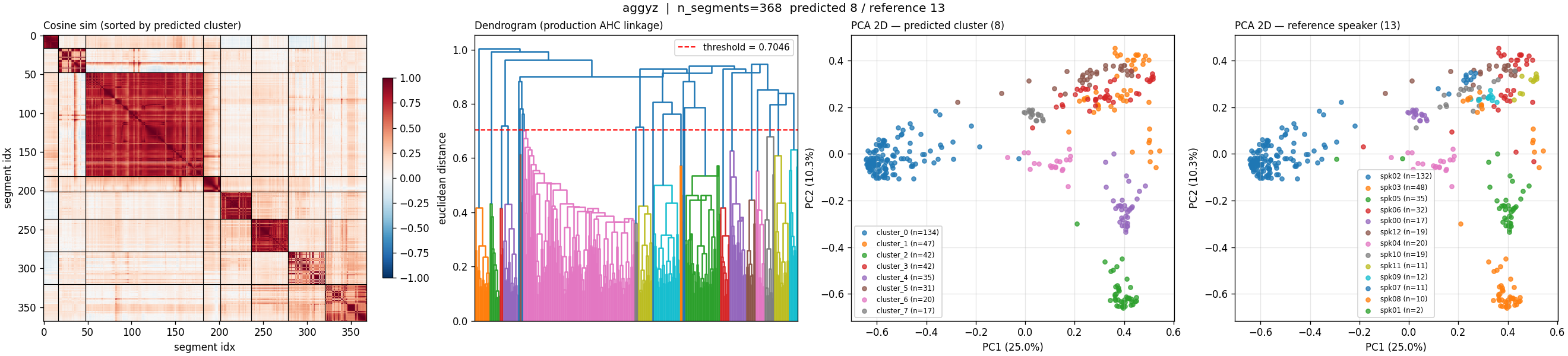}\\[2pt]
\includegraphics[width=\textwidth]{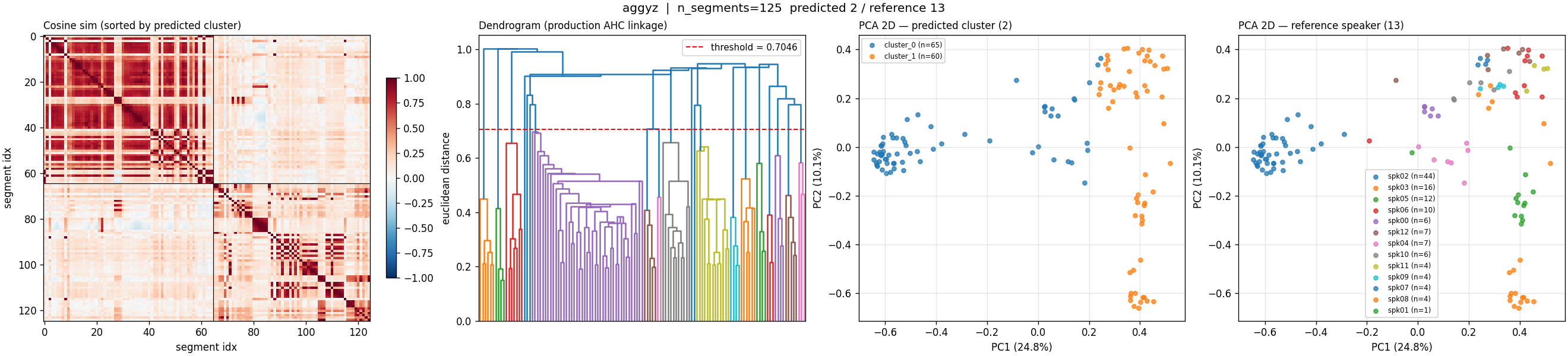}
\caption{Clustering intermediates for VoxConverse \texttt{aggyz} (13 reference
speakers). Top: stride~1 baseline (8 predicted clusters, 368 embeddings).
Bottom: stride~3 (2 predicted clusters, 125 embeddings). Left to right: cosine
similarity matrix (sorted by predicted cluster), AHC dendrogram with the
production threshold, PCA of embeddings colored by predicted cluster, and by
reference speaker. The coarser stride collapses the similarity structure and
over-merges speakers.}
\label{fig:cluster}
\end{figure*}

We confirm this mechanism by visualizing the clustering intermediates on
VoxConverse files where the degradation is largest
(Fig.~\ref{fig:cluster}). At stride~1, the cosine-similarity matrix shows clear
speaker blocks and the dendrogram has many branches above the threshold; at
stride~3, the number of segment embeddings falls to about one third (e.g.\ on
\texttt{aggyz}, $368\!\to\!125$), the similarity structure collapses into a few
blurred blocks, and the predicted speaker count drops far below the reference
(\texttt{aggyz} 13~ref: $8\!\to\!2$; \texttt{qeejz} 14~ref: $11\!\to\!4$;
\texttt{gukoa} 10~ref: $6\!\to\!2$). Cluster purity falls correspondingly
(\texttt{aggyz} $0.85\!\to\!0.48$), confirming over-merging rather than a
segmentation failure.

\paragraph{A rejected alternative: file-adaptive stride.}
We also tested choosing the stride per file. An oracle (best stride per file)
recovers DER but yields only $1.10\times$ speedup on VoxConverse, because 83\% of
files require stride~1. Turn density is uncorrelated with the DER change
($\rho=-0.04$). We therefore do not pursue dynamic stride; the speedup comes from
stride and per-chunk embedding, and the DER recovery from relative mcs.

\section{Relative Minimum Cluster Size}
\label{sec:relative}

Setting $\mathrm{mcs}=\mathrm{round}(f\cdot n)$ with $f=0.01$ recovers
VoxConverse DER from $0.113$ to $0.079$, about 89\% of the lost accuracy, while
AMI DER stays flat at $0.083$ (Table~\ref{tab:main}). A single value of $f$ thus
works across both datasets.

\begin{figure}[t]
\centering
\includegraphics[width=0.85\columnwidth]{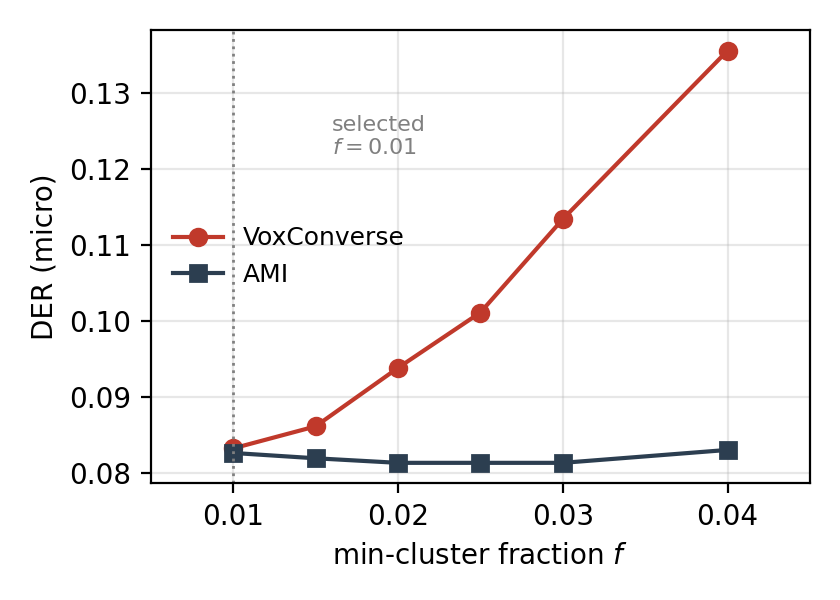}
\caption{DER vs.\ the min-cluster fraction $f$ (stride3 + per-chunk, MPS) on the
probe set. AMI is flat; VoxConverse degrades monotonically with $f$. The two are
jointly best at $f=0.01$.}
\label{fig:fsweep}
\end{figure}

We chose $f$ by sweeping it on a smaller probe set---the full AMI test set and a
speaker-stratified 39-file VoxConverse subset---and measuring DER directly
(Fig.~\ref{fig:fsweep}). AMI is insensitive to $f$ over $0.01$--$0.03$
(DER $0.081$--$0.083$), consistent with its abundant embeddings, whereas
VoxConverse degrades monotonically as $f$ grows, because a larger fraction
revives the over-merging that the relative rule is meant to avoid. The two curves
are jointly best at the low end, so we take $f=0.01$ and confirm it on the full
sets (the subset DER differs slightly from the full-set values in
Table~\ref{tab:main}; e.g.\ VoxConverse $0.083$ on the subset vs.\ $0.079$ on the
full test set).

We note this is a re-parameterization of an existing
term rather than a new operator: Pyannote's AHC already contains a relative rule
($\min(12,\mathrm{round}(0.1\cdot n))$), but with coefficient $0.1$ and a cap of
12 both datasets saturate at the cap; the key is a smaller coefficient
($\sim$0.01) and removing the cap. The recovery is not complete---a residual
$+0.004$ over the stride-1 baseline remains, attributable to the reduced
segmentation resolution.

\section{Cross-Dataset Validation: MSDWild}
\label{sec:msdwild}

On MSDWild (many.val, 177 files), the accelerated pipeline reduces RTF from
$0.0045$ to $0.0013$ on CUDA (about $3.5\times$). DER changes from $0.384/0.472$
to $0.389/0.484$: the micro DER change stays within 0.01, but the macro DER
change is $+0.012$, slightly exceeding it (Table~\ref{tab:msdwild}). On this hard
dataset the relative-mcs scheme and a fixed $\mathrm{mcs}=12$ perform almost
identically (0.389/0.484 vs.\ 0.393/0.481; Table~\ref{tab:msdwild}), so the
cluster-size scheme provides little benefit here. This contrasts with
VoxConverse, where the same two settings differ by more than 0.03 DER, and is
consistent with our diagnosis: MSDWild's degradation is not dominated by the
embedding-budget mechanism that relative mcs addresses.

\begin{table}[t]
\centering
\caption{MSDWild many.val (DER with skip\_overlap=true; CUDA).}
\label{tab:msdwild}
\setlength{\tabcolsep}{4pt}
\begin{tabular}{lcc}
\toprule
Configuration & DER micro/macro & RTF \\
\midrule
baseline (stride1, mcs12) & 0.384 / 0.472 & 0.0045 \\
stride3 + per-chunk, mcs12 & 0.393 / 0.481 & 0.0012 \\
stride3 + per-chunk + relative mcs & 0.389 / 0.484 & \textbf{0.0013} \\
\bottomrule
\end{tabular}
\end{table}

To place all three datasets on the same device, we re-ran the accelerated
configuration on MSDWild on MPS. The DER agrees with the CUDA run to within
numerical noise ($0.388/0.484$ vs $0.389/0.484$); the small difference arises
because float16 embeddings differ slightly between backends, flipping a few
borderline clustering decisions whose effects largely cancel in aggregate. The
RTF ($0.007$) is comparable to AMI and VoxConverse on the same hardware
(Table~\ref{tab:crossdevice}). The final pipeline thus runs at a similar,
small RTF ($0.005$--$0.007$ on an Apple~M4 laptop) across all three datasets.
At this rate even a two-hour recording is diarized in under a minute (roughly
$40$--$50$\,s), turning a task one would start and walk away from into one a user
can sit through in place.

\begin{table}[t]
\centering
\caption{Final configuration (stride3 + per-chunk + relative mcs, $f{=}0.01$)
across datasets. AMI/MSDWild on MPS (Apple~M4); VoxConverse on CUDA (RTX~5070~Ti).}
\label{tab:crossdevice}
\begin{tabular}{lcc}
\toprule
Dataset & DER micro/macro & RTF \\
\midrule
AMI test & 0.083 / 0.088 & 0.005 (MPS) \\
VoxConverse test & 0.079 / 0.086 & 0.00083 (CUDA) \\
MSDWild many.val & 0.388 / 0.484 & 0.007 (MPS) \\
\bottomrule
\end{tabular}
\end{table}

\section{Limitations}
\label{sec:limitations}

MSDWild DER is high in absolute terms ($\sim$0.44), and the cluster-size scheme
does not help there; the ``DER preserved'' claim is dataset-dependent. The
VoxConverse recovery is partial ($+0.004$ residual). The fraction $f=0.01$ was
selected by sweeping DER on the test sets (the full AMI test set and a
speaker-stratified VoxConverse subset), without a held-out development split;
the value should thus be read as a robust setting rather than a tuned optimum,
and a proper dev/test separation is left for future work. Evaluation conditions
are not fully uniform across datasets (MSDWild is more naturally evaluated
without skip\_overlap); we report conditions explicitly. RTF measurements mix MPS
and CUDA and should be read per device.

\section{Conclusion}
\label{sec:conclusion}

A coarse-stride, per-chunk pipeline accelerates speaker diarization several-fold
but breaks on in-the-wild data due to speaker under-counting. Making the minimum
cluster size relative to the embedding budget resolves this with a single
hyperparameter, preserving DER on AMI and recovering most of the loss on
VoxConverse; on the harder MSDWild the gain is limited. The accelerated pipeline
reaches up to $12.2\times$ speedup on AMI (MPS) over our CAM++ baseline on
consumer hardware.

\end{document}